\documentclass[11pt]{article}
\usepackage{graphicx}

\newcommand{\BABARPubYear}    {04}

\newcommand{\BABARConfNumber} {23}
\newcommand{\SLACPubNumber} {10645}

\input pubboard/babarsym
\RequirePackage{xspace}

\newcommand{\pvec}{{\bf p}}

\newcommand{\acp}{\ensuremath{\calA_{ch}}}
\newcommand{\calB}{\ensuremath{{\cal B}}}

\newcommand{\DE}{\ensuremath{\Delta E}}

\newcommand{\mres}{\ensuremath{m_{\rm res}}}
\newcommand{\xf}{\ensuremath{{\cal F}}}
\newcommand{\hel}{\ensuremath{{\cal H}}}
\newcommand{\thetaT}{\ensuremath{\theta_{\rm T}}}
\newcommand{\costhr}{\ensuremath{\cos\thetaT}}

\newcommand\etal{{\it et al.}}
\newcommand{\half}{\ensuremath{{1\over2}}}

\newcommand{\bma}[1]{\boldmath{$#1$}}

\newcommand{\bfig}{\begin{figure}[htbpc!]}
\newcommand{\efig}{\end{figure}}
\newcommand\bef{\begin{figure}}
\newcommand\edf{\end{figure}}
\newcommand\dbline{\noalign{\vskip 0.10truecm\hrule}\noalign{\vskip 2pt}\noalign{\hrule\vskip 0.10truecm}}
\providecommand{\tbline}{\noalign{\vskip 0.05truecm\hrule\vskip0.05truecm}}
\newcommand\sgline{\noalign{\vskip 0.10truecm\hrule\vskip 0.10truecm}}
\newcommand\beq{\begin{equation}}
\newcommand\eeq{\end{equation}}
\newcommand\bear{\begin{array}}
\newcommand\enar{\end{array}}
\newcommand\beqa{\begin{eqnarray}}
\newcommand\eeqa{\end{eqnarray}}
\newcommand\ben{\begin{enumerate}}
\newcommand\een{\end{enumerate}}

\newcommand{\UfourS}{\ensuremath{\Upsilon(4S)}}

\newcommand{\etagg}{\ensuremath{\eta_{\gaga}}}
\newcommand{\etappp}{\ensuremath{\eta_{3\pi}}}
\newcommand{\etatogg}{\ensuremath{\eta\ra\gaga}}
\newcommand{\etatoppp}{\ensuremath{\eta\ra\pi^+\pi^-\pi^0}}

\newcommand{\etapepp}{\ensuremath{\etapr_{\eta\pi\pi}}}
\newcommand{\etaptoepp}{\ensuremath{\etapr\ra\eta\pip\pim}}
\newcommand{\etaprg}{\ensuremath{\etapr_{\rho\gamma}}}
\newcommand{\etaptorg}{\ensuremath{\etapr\ra\rho^0\gamma}}

\newcommand{\omtoppp}{\ensuremath{{\omega\ra\pip\pim\piz}}}

\newcommand{\kzs}{\ensuremath{\KS}}

\newcommand{\hp}{\ensuremath{h^{+}}}

\newcommand{\fetapip}{\ensuremath{\eta\pi^+}}
\newcommand{\etapip}{\ensuremath{\Bp\ra\fetapip}}

\newcommand{\fetaKp}{\ensuremath{\eta K^+}}
\newcommand{\etaKp}{\ensuremath{\Bp\ra\fetaKp}}

\newcommand{\fetaKz}{\ensuremath{\eta\Kz}}
\newcommand{\etaKz}{\ensuremath{\Bz\ra\fetaKz}}
\newcommand{\BetaKz}{\ensuremath{\calB(\etaKz)}}
\newcommand{\retaKz}{\ensuremath{xx^{+xx}_{-xx}\pm xx}}
\newcommand{\RetaKz}{\ensuremath{(\retaKz)\times 10^{-6}}}

\newcommand{\setaKz}{\ensuremath{xx}}

   \newcommand{\fetaggKz}{\ensuremath{\eta_{\gaga}\Kz}}
   \newcommand{\etaggKz}{\ensuremath{\Bz\ra\fetaggKz}}

   \newcommand{\fetapppKz}{\ensuremath{\eta_{3\pi}\Kz}}
   \newcommand{\etapppKz}{\ensuremath{\Bz\ra\fetapppKz}}

\newcommand{\fetaomega}{\ensuremath{\eta\omega}\xspace}
\newcommand{\etaomega}{\ensuremath{\Bz\ra\fetaomega}\xspace}
\newcommand{\Betaomega}{\ensuremath{\calB(\etaomega)}\xspace}
\newcommand{\retaomega}{\ensuremath{xx^{+xx}_{-xx}\pm xx}\xspace}
\newcommand{\Retaomega}{\ensuremath{(\retaomega)\times 10^{-6}}\xspace}
\newcommand{\uletaomega}{\ensuremath{xx}\xspace}
\newcommand{\ULetaomega}{\ensuremath{\uletaomega\times 10^{-6}}\xspace}
\newcommand{\setaomega}{\ensuremath{xx}\xspace}
   \newcommand{\fetaggomega}{\ensuremath{\eta_{\gaga} \omega}\xspace}
   \newcommand{\etaggomega}{\ensuremath{\Bz\ra\fetaggomega}\xspace}
   \newcommand{\fetapppomega}{\ensuremath{\eta_{3\pi} \omega}\xspace}
   \newcommand{\etapppomega}{\ensuremath{\Bz\ra\fetapppomega}\xspace}

\newcommand{\fetarhop}{\ensuremath{\eta\rho^+}}
\newcommand{\etarhop}{\ensuremath{\Bp\ra\fetarhop}}
\newcommand{\Betarhop}{\ensuremath{\calB(\etarhop)}}
\newcommand{\retarhop}{\ensuremath{xx^{+xx}_{-xx}\pm xx}}
\newcommand{\Retarhop}{\ensuremath{(\retarhop)\times 10^{-6}}}

\newcommand{\Aetarhop}{\ensuremath{xx\pm xx \pm xx}}
\newcommand{\setarhop}{\ensuremath{xx}}
  \newcommand{\fetaggrhop}{\ensuremath{\eta_{\gamma\gamma} \rho^+}}
  \newcommand{\etaggrhop}{\ensuremath{\Bp\ra\fetaggrhop}}
  \newcommand{\fetappprhop}{\ensuremath{\eta_{3\pi} \rho^+}}
  \newcommand{\etappprhop}{\ensuremath{\Bp\ra\fetappprhop}}

\newcommand{\fetapK}{\ensuremath{\etapr K}}
\newcommand{\etapK}{\ensuremath{\B\ra\fetapK}}

\newcommand{\fetaph}{\ensuremath{\etapr\hp}}
\newcommand{\etaph}{\ensuremath{\Bp\ra\fetaph}}

\newcommand{\fetappip}{\ensuremath{\etapr\pip}}
\newcommand{\etappip}{\ensuremath{\Bp\ra\fetappip}}
\newcommand{\Betappip}{\ensuremath{\calB(\Bp\ra\etapr \pip)}}
\newcommand{\retappip}{\ensuremath{xx^{+xx}_{-xx} \pm xx}}
\newcommand{\Retappip}{\ensuremath{(\retappip)\times 10^{-6}}}

\newcommand{\Aetappip}{\ensuremath{xx\pm xx}}
\newcommand{\setappip}{\ensuremath{xx}}
   \newcommand{\fetapepppip}{\ensuremath{\etapr_{\eta\pi\pi} \pi^+}}
   \newcommand{\etapepppip}{\ensuremath{\Bp\ra\fetapepppip}}
   \newcommand{\fetaprgpip}{\ensuremath{\etapr_{\rho\gamma} \pi^+}}
   \newcommand{\etaprgpip}{\ensuremath{\Bp\ra\fetaprgpip}}

\newcommand{\fetapKp}{\ensuremath{\etapr K^+}}
\newcommand{\etapKp}{\ensuremath{\Bp\ra\fetapKp}}

\renewcommand{\retaomega}{\ensuremath{1.2\pm 0.6\pm 0.2}}
\renewcommand{\uletaomega}{\ensuremath{2.3}}
\renewcommand{\setaomega}{\ensuremath{2.2}}

\renewcommand{\retaKz}{\ensuremath{2.5\pm0.8\pm 0.1}}

\renewcommand{\setaKz}{\ensuremath{4.2}}

\renewcommand{\retarhop}{\ensuremath{8.6\pm 2.2\pm 1.1}}
\renewcommand{\Aetarhop}{\ensuremath{7\pm 19 \pm 2}}
\renewcommand{\setarhop}{\ensuremath{4.2}}

\renewcommand{\retappip}{\ensuremath{4.2\pm 1.0\pm 0.5}}

\renewcommand{\Aetappip}{\ensuremath{24\pm19\pm 1}}
\renewcommand{\setappip}{\ensuremath{4.8}}

\long\def\inst#1{\par\nobreak\kern 4pt\nobreak
    {\it #1}\par\vskip 10pt plus 3pt minus 3pt}

\begin{document}
{\pagestyle{empty}

\begin{flushright}
\babar-CONF-\BABARPubYear/\BABARConfNumber \\
SLAC-PUB-\SLACPubNumber \\
August 2004 \\
\end{flushright}

\par\vskip 5cm

\begin{center}
\Large\boldmath\bf Searches for Charmless Decays \etaomega, \etaKz, \etarhop, and \etappip
\end{center}
\bigskip

\begin{center}
\large The \babar\ Collaboration\\
\mbox{ }\\
\today
\end{center}
\bigskip \bigskip

\begin{center}
\large \bf Abstract
\end{center}
We present measurements of branching fractions
for four previously unobserved $B$-meson decays with an $\eta$ or
\etapr\ meson in the final 
state.  The data sample corresponds to 182 million \BB\ pairs produced
from \epem\ annihilation at the \UfourS\ resonance.  We measure the
following branching fractions in units of $10^{-6}$:
$\Betaomega=\retaomega$ ($<\uletaomega$ at 90\% C.L.), 
$\BetaKz=\retaKz$, 
$\Betarhop=\retarhop$, 
and 
$\Betappip=\retappip$, where the first error quoted is statistical and
the second systematic. 
The charge asymmetries are 
$\acp(\etarhop)=(\Aetarhop)\%$ and 
$\acp(\etappip)=(\Aetappip)\%$.  
All results are preliminary.

\vfill
\begin{center}

Submitted to the 32$^{\rm nd}$ International Conference on High-Energy Physics, ICHEP 04,\\
16 August---22 August 2004, Beijing, China

\end{center}

\vspace{1.0cm}
\begin{center}
{\em Stanford Linear Accelerator Center, Stanford University, 
Stanford, CA 94309} \\ \vspace{0.1cm}\hrule\vspace{0.1cm}
Work supported in part by Department of Energy contract DE-AC03-76SF00515.
\end{center}

\newpage
} 

%
\begin{center}
\small

The \babar\ Collaboration,
\bigskip

%
B.~Aubert,
R.~Barate,
D.~Boutigny,
F.~Couderc,
J.-M.~Gaillard,
A.~Hicheur,
Y.~Karyotakis,
J.~P.~Lees,
V.~Tisserand,
A.~Zghiche
\inst{Laboratoire de Physique des Particules, F-74941 Annecy-le-Vieux, France }
% ...

\end{center}\newpage

\section{INTRODUCTION}
\label{sec:Introduction}

We report results for measurements of the decay branching fractions of
\Bz\ to the charmless final states\footnote{Except as noted explicitly,
we use a particle name to denote either member of a charge conjugate
pair.} \fetaomega\ and \fetaKz, and of \Bp\ to \fetarhop\ and
\fetappip.  None of these decays have been observed definitively
\cite{isoscalarPRL, etappiPRL, etaPRD%
}.  Measurements of the related decays \etaKp, \etapip,
and \etapK\ were published recently \cite{etappiPRL,etaprPRL}.
Charmless decays with kaons are usually expected to be dominated by
$b\ra s$ loop (``penguin") transitions, while $b\ra u$ tree transitions
are typically larger for the decays with pions and $\rho$ mesons.
However the $\B\ra\eta K$ decays are especially 
interesting since they are suppressed relative to the abundant $\B\ra\etapr K$ 
decays due to destructive interference between two penguin amplitudes
\cite{Lipkin}.
The CKM-suppressed $b\ra u$ amplitudes may interfere significantly with
penguin amplitudes, possibly leading to large direct \CP\ violation in
\etarhop\ and \etappip \cite{directCP}; numerical estimates are
available in a few cases \cite{acpgrabbag,acpQCDfact}.  We search for
such direct  
\CP\ violation by measuring the charge asymmetry $\acp \equiv
(\Gamma^--\Gamma^+)/(\Gamma^-+\Gamma^+)$ in the rates
$\Gamma^\pm=\Gamma(B^\pm\ra f^\pm)$, for each observed charged final
state $f^\pm$.

Charmless $B$ decays are becoming useful to test the
accuracy of theoretical predictions
\cite{acpQCDfact,FU,FUglob,ALI,LEPAGE,BENEKE,BN,chiang,chiangGlob}. 
Phenomenological fits to the branching fractions and charge asymmetries
can be used to understand the importance of tree and penguin contributions 
and may provide sensitivity to the CKM angle $\gamma$ \cite{chiangGlob}.

\section{THE \babar\ DETECTOR AND DATASET}
\label{sec:babar}

The results presented here are based on data collected with the \babar\
detector~\cite{BABARNIM} at the PEP-II asymmetric \epem\
collider~\cite{pep} located at the Stanford Linear Accelerator Center.
The samples come from an integrated luminosity of 166~fb$^{-1}$ recorded
at the $\Upsilon (4S)$ resonance (center-of-mass energy $\sqrt{s}=10.58\
\gev$).  This corresponds to $182\pm2$ million \BB\ pairs.

Charged particles from the \epem\ interactions are detected and their
momenta measured by a combination of a vertex tracker (SVT) consisting
of five layers of double-sided silicon microstrip detectors and a
40-layer central drift chamber, both operating in the 1.5-T magnetic
field of a superconducting solenoid. We identify photons and electrons 
using a CsI(Tl) electromagnetic calorimeter (EMC).
Further charged particle identification (PID) is provided by the average energy
loss ($dE/dx$) in the tracking devices and by an internally reflecting
ring imaging Cherenkov detector (DIRC) covering the central region.

\section{SAMPLE SELECTION}
\label{sec:EvSel}

\begin{table}[btp]
\begin{center}
\caption{
Selection requirements on the invariant mass of $B$ daughter resonances
(in \mev).
}
\label{tab:rescuts}
\begin{tabular}{lc}
\dbline
State		& Requirement			\\
\sgline						
\etagg		&$490< m(\gaga)<600$		\\
\etappp		&$520<m(\pi\pi\pi)< 570$	\\
\etapepp	&$910<m(\eta\pi\pi)<1000$	\\
\etaprg		&$910<m(\rho\gamma)<1000$	\\
$\omega$	&$735<m(\pi\pi\pi)<825$		\\
$\rho^+$	&$470 < m(\pi\pi) <1070$	\\
$\rho^0$	&$510 < m(\pi\pi) <1060$	\\
\piz		&$120 < m(\gaga) <150$		\\
\kzs		&$486 < m(\pi\pi) < 510$	\\
\dbline
\end{tabular}
\vspace{-5mm}
\end{center}
\end{table}

We reconstruct $\eta$, $\etapr$, $\omega$, $\rho^+$, $\rho^0$, \piz, and
\KS\ candidates through their decays \etatogg\ (\etagg), \etatoppp\
(\etappp), \etaptoepp\ (\etapepp), \etaptorg\ (\etaprg),
\omtoppp, $\rho^+\ra\pip\piz$, $\rho^0\ra\pip\pim$, $\piz\ra\gaga$, and
$\kzs\ra\pip\pim$.  
We make the requirements given in Table \ref{tab:rescuts}\ on the invariant
mass of these particles' final states.
For the $\eta$, $\omega$, and \etapr\ invariant masses these
requirements are set loose enough to include sidebands, as these mass
values are treated as observables in the maximum-likelihood (ML) fit
described below.
For \kzs\ candidates we further require
the three-dimensional flight distance from the beam spot to be 
greater than three times its uncertainty in a fit that requires
consistency between the flight and momentum directions.  For modes with
$B\ra\eta\ra\gamma\gamma$ we impose a mode-dependent requirement on the
decay angle to reject backgrounds reconstructed as very asymmetric
decays. 

We make several PID requirements to ensure the identity of the charged
pions and 
kaons.  Secondary pions in \etappp, \etapr, and $\omega$ candidates
are rejected if their DIRC, $dE/dx$, and EMC outputs satisfy tight
consistency with kaons, protons, or electrons.  For the
\Bp\ decays to an $\etapr$ meson and a charged pion or kaon,
the latter (primary) track must have an associated DIRC signal with a
Cherenkov angle within $3.5$ standard deviations ($\sigma$) of the
expected value for either the $\pi$ or $K$ hypothesis.  The discrimination
between pion and kaon primary tracks is treated in the ML fit.

The number of candidates found per event is at or below about 1.10 for
all modes except for those with the final states $\etappp\rho$ and
$\etappp\omega$ where it is about 1.3.  We choose the candidate whose
daughter resonance mass(es) lie nearest the expected mean value.

A $B$-meson candidate is characterized kinematically by the
energy-substituted mass $\mes=\lbrack{(\half
s+\pvec_0\cdot\pvec_B)^2/E_0^2-\pvec_B^2}\rbrack^\half$ and energy
difference $\DE = E_B^*-\half\sqrt{s}$, where the subscripts $0$ and $B$
refer to the initial \UfourS\ and to the $B$ candidate, respectively,
and the asterisk denotes the \UfourS\ frame. The resolution on \DE\
(\mes) is about 30 MeV ($3.0\ \mev$). We require $|\DE|\le0.2$ GeV and
$5.25\le\mes\le5.29\ \gev$, and include both of these observables in the
ML fit.

\section{BACKGROUNDS}
\label{sec:Bkg}
Backgrounds arise primarily from random combinations in continuum
$\epem\ra\qqbar$ events ($q=u,d,s,c$).  We reject these by using the angle
\thetaT\ in the \UfourS\
frame between the thrust axis of the $B$ candidate and that of the rest
of the charged tracks and neutral calorimeter clusters in 
the event.  The distribution of $|\costhr|$ is sharply peaked near $1.0$
for combinations drawn from jet-like \qqbar\ pairs, and nearly uniform
for $B$-meson pairs.  We require $|\costhr|<0.9$ for all modes except
the high-background \etaprgpip\ decay.  For this mode we determined that the
sensitivity is maximal with $|\costhr|<0.65$, based on the expected
signal yield and its background-dominated statistical error.  In the ML 
fit we also use a Fisher discriminant \xf\ \cite{fisher}\ that combines
four variables defined in the \UfourS\ frame: the angles 
with respect to the beam axis of the $B$ momentum and $B$ thrust axis,
and the zeroth and second angular moments 
$L_{0,2}$ of the energy flow about the $B$ thrust axis.  The moments are
defined by $ L_j = \sum_i p_i\times\left|\cos\theta_i\right|^j,$ where
$\theta_i$ is the angle with respect to the $B$ thrust axis of track or
neutral cluster $i$, $p_i$ is its momentum, and the sum excludes the $B$
candidate daughters.

For the \etatogg\ modes we use additional event-selection criteria to
reduce \BB\ backgrounds from several charmless final states. We reduce
background from $B\ra\pip\piz$, $\Kp\piz$, and $\Kz\piz$ by rejecting
\etagg\ candidates that share a photon with any \piz\ candidate having
momentum between 1.9 and 3.1 GeV/c in the \UfourS\ frame.  Additionally,
for \etaKz\ we require $E^*_{\gamma} < 2.4$ GeV to suppress background from
$B\ra\Kstar\gamma$ and related radiative-penguin decays.  

We use Monte Carlo (MC) simulation \cite{geant} for an initial estimate
of the residual charmless \BB\ background.  Most of the contribution
from $b\ra c$ decays has a dependence on the ML fit observables that is
similar to that for continuum events, and thus can be modeled as part of
the continuum component.  With a 
survey from MC we identify the few (mostly charmless) decays that may
survive the candidate selection.  We find these contributions to be
negligible for several of our modes.  Where they are not we include a
component in the ML fit to account for them.

\section{MAXIMUM LIKELIHOOD FIT}
\label{sec:MLfit}
We obtain yields and \acp\ from extended unbinned 
maximum-likelihood fits with input observables \DE, \mes, \xf, and
$\mres$ (the mass of the $\eta$, \etapr, $\rho^+$, or $\omega$ candidate).
For the $\omega$ decays we also use $\hel\equiv |\cos{\theta_H}|$, and for charged modes 
the PID variable $S_{\pi,K}$.  The helicity angle $\theta_H$ is 
defined as the angle, measured in the $\omega$ rest frame, between the normal 
to the $\omega$ decay plane and the flight direction of the $\omega$
with respect to its parent $B$.  
We incorporate PID information by using $S_\pi$ ($S_K$), 
the number of standard deviations between the measured Cherenkov angle
and that expected for pions (kaons).

For each event $i$, hypothesis $j$ (signal, continuum background, 
\BB\ background), and flavor (primary \pip\ or \Kp)
$k$, we define the  probability density function (PDF)
\begin{eqnarray}
{\cal P}^i_{jk} &=&  {\cal P}_j (\mes^i) {\cal  P}_j (\DE^i_k) 
 { \cal P}_j(\xf^i) {\cal P}_j (\mres^i) \nonumber \\
 && \times\left[{\cal P}_j
(S^i_k)\right]\left[{\cal P}_j (\hel^i)\right].
\end{eqnarray}
The terms in brackets for $S$ and \hel\ pertain to modes with a primary charged
track or $\omega$ daughters, respectively.  The absence of correlations
among observables in the background ${\cal P}^i_{jk}$ is confirmed in the
(background-dominated) data samples entering the fit.  For the signal
component, we correct for the effect of the neglect of small correlations 
(see below).  The likelihood function is
\begin{equation}
{\cal L} = \exp{(-\sum_{j,k} Y_{jk})}
\prod_i^{N}\left[\sum_{j,k} Y_{jk} {\cal P}^i_{jk}\right]\,,
\end{equation}
where $Y_{jk}$ is the yield of events of hypothesis $j$ and flavor $k$
found by maximizing \calL, and $N$ is the number of events in the sample.  
  
For the signal and \BB\ background components we determine the PDF
parameters from simulation.  For the continuum background we use
(\mes,\,\DE) sideband data to obtain initial values, before applying the
fit to data in the signal region, and ultimately by leaving them free in
the final fit.  We parameterize each of the functions ${\cal P}_{\rm
sig}(\mes),\ {\cal P}_{\rm sig}(\DE_k),\ { \cal P}_j(\xf),\ { \cal
P}_j(S_k)$ and the peaking components of ${\cal P}_j(\mres)$ with either
a Gaussian, the sum of two Gaussians or an asymmetric Gaussian function
as required to describe the distribution.  Slowly varying distributions
(mass, energy or helicity-angle for combinatorial background) are
represented by linear or quadratic dependencies.  The peaking and
combinatorial components of the $\omega$ mass spectrum each have their
own $\hel$ shapes.  The combinatorial background in \mes\ is described by
the function $x\sqrt{1-x^2}\exp{\left[-\xi(1-x^2)\right]}$, with
$x\equiv2\mes/\sqrt{s}$ and parameter $\xi$.  Large control samples of
$B$ decays to charmed final states of similar topology are used to
verify the simulated resolutions in \DE\ and \mes.  Where the control
data samples reveal differences from MC in mass or energy offset or
resolution, we shift or scale the resolution function used in the likelihood
fits.

Before applying the fitting procedure to the data to extract the signal
yields we subject it to several tests.  Internal consistency is checked
with fits to ensembles of ``experiments" generated by Monte Carlo from
the PDFs.  From these we establish the number of parameters associated
with the PDF shapes that can be left free in addition to the
yields.  Ensemble distributions of the fitted parameters verify that 
the generated values are reproduced with the expected resolution.  The
ensemble distribution of $\ln{\calL}$ itself provides a reference to
check the goodness of fit of the final measurement once it has been
performed. 

We evaluate possible biases from our neglect of correlations among 
discriminating variables in the PDFs by fitting ensembles of simulated
experiments into which we have embedded the expected
number of signal events randomly extracted from the fully simulated MC
samples.  We find 
a positive bias of $\lsim 1$ event for the \fetaomega\ and \fetaKz\
modes.  For \fetarhop\ and \fetappip\ it is 7 to 17 events
($\sim15$\%\ of the yield).
Events from a weighted
mixture of simulated \BB\ background decays are included where
significant, and so the bias we measure includes the effect of crossfeed
from these modes. 

\section{FIT RESULTS}
\label{sec:Results}

Free parameters of the fit include signal and background yields,
background PDF parameters, and for charged modes the signal and
background \acp.  The free background PDF parameters are mean, width,
and skewness
for \xf, slope for \DE, slope of the combinatorial component and peak
fraction for resonance mass, and $\xi$ for \mes.

The branching fraction for each decay chain is obtained from
\beq
\calB = \frac{Y-Y_b}{\epsilon\prod{\calB_i}N_B}\,,
\eeq
where $Y$ is the yield of signal events from the fit, $Y_b$ is the fit
bias discussed in the previous section, $\epsilon$ is the efficiency,
$\prod{\calB_i}$ is the product of daughter branching fractions that
were forced to unity in the determination of $\epsilon$, and $N_B$ is
the number of produced \Bz\ or \Bp\ mesons.  In Table \ref{tab:results} we
show for each decay mode the measured branching fraction together with
the event yields and efficiencies.
We assume that the decay rates 
of the \UfourS\ to \BpBm\ and \BzBzb\ are equal.
The estimated purity is the ratio of the signal yield to the effective
background plus signal; the sum of effective bkg plus signal is
represented by the square of the uncertainty of the signal yield.

\begin{figure}[!tb]
\vspace{0.5cm}
\includegraphics[bb=275 40 556 761,clip,angle=270,width=1.0\linewidth]{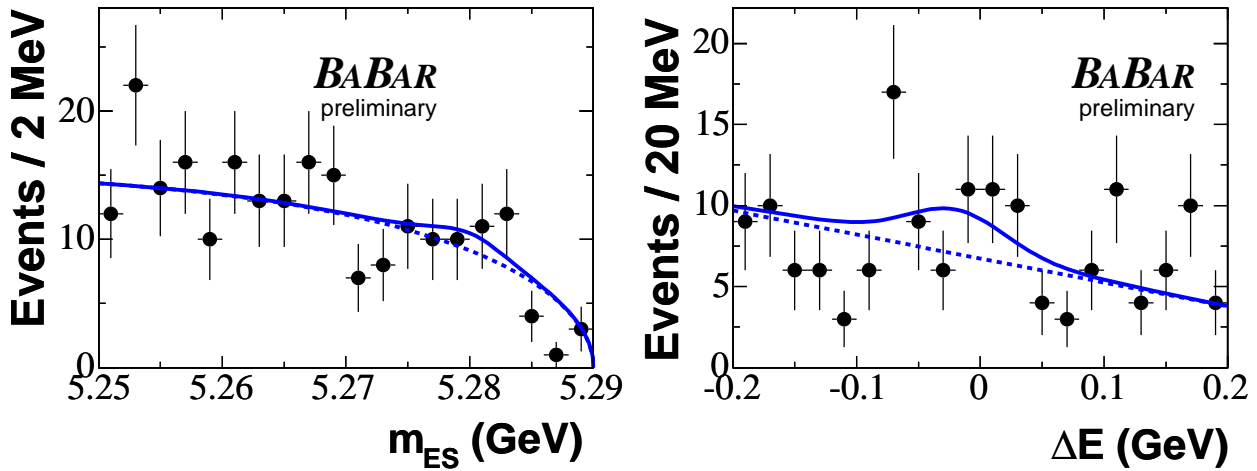}
 \caption{\label{fig:proj_etaomega}
Projections of the $B$ candidate \mes\ (left) and \DE\ (right) for \etaomega.
Points with errors represent data,  
solid curves the full fit functions,
 and dashed curves the background functions.
These plots are made with a requirement on the likelihood and thus do not 
show all events in the data sample.
  }
\end{figure}

\begin{figure}[!tb]
\vspace{0.5cm}
\includegraphics[bb=275 40 556 761,clip,angle=270,width=1.0\linewidth]{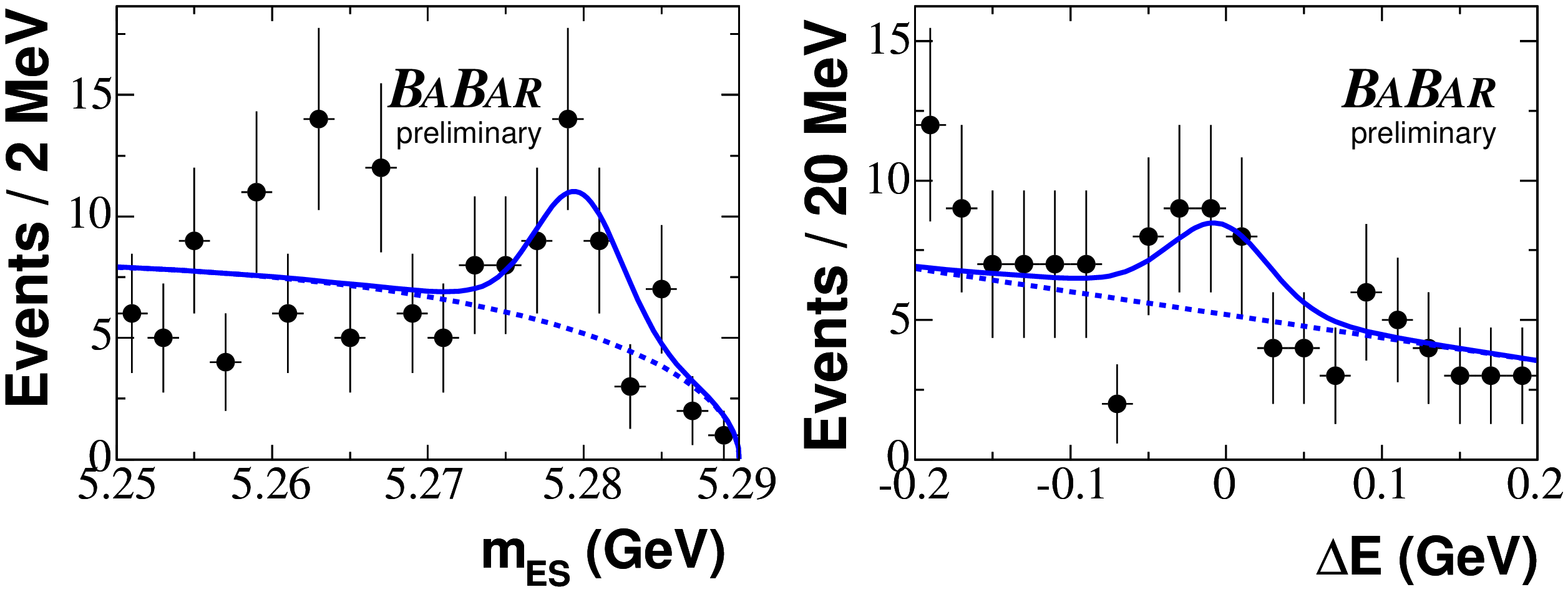}
 \caption{\label{fig:proj_etaKz}
Projections of the $B$ candidate \mes\ (left) and \DE\ (right) for \etaKz.
Points with errors represent data,  
solid curves the full fit functions,
 and dashed curves the background functions.
These plots are made with a requirement on the likelihood and thus do not 
show all events in the data sample.
  }
\end{figure}

\begin{figure}[!tb]
\vspace{0.5cm}
\includegraphics[angle=0,width=\linewidth]{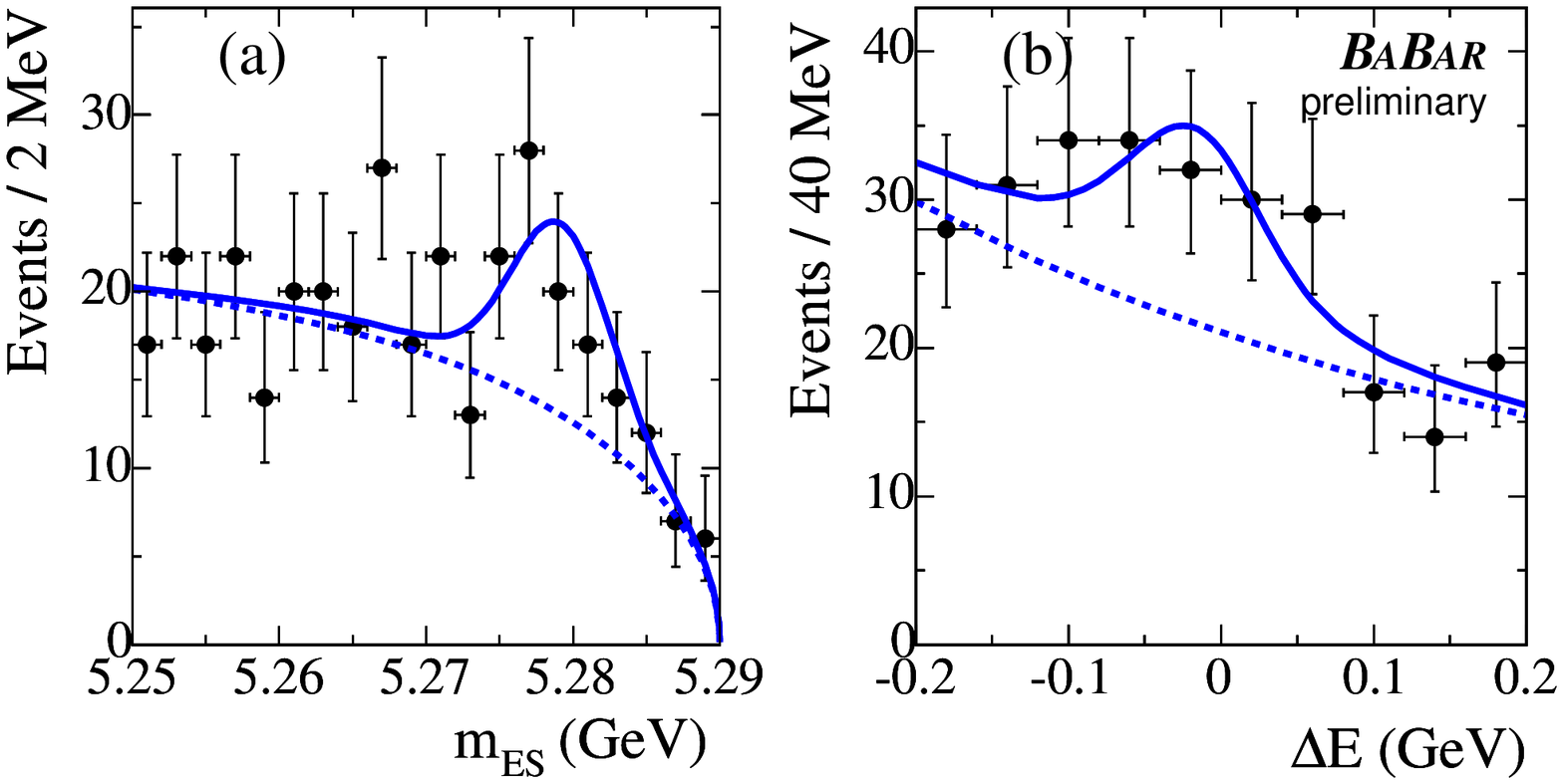}
 \caption{\label{fig:proj_deMb_etarhop}
Projections of the $B$ candidate \mes\ (a) and \DE\ (b) for \etarhop.
Points with errors represent data,  
solid curves the full fit functions,
 and dashed curves the background functions.
These plots are made with a requirement on the likelihood and thus do not 
show all events in the data sample.
  }
\end{figure}

\begin{figure}[!tb]
\vspace{0.5cm}
\includegraphics[angle=0,width=\linewidth]{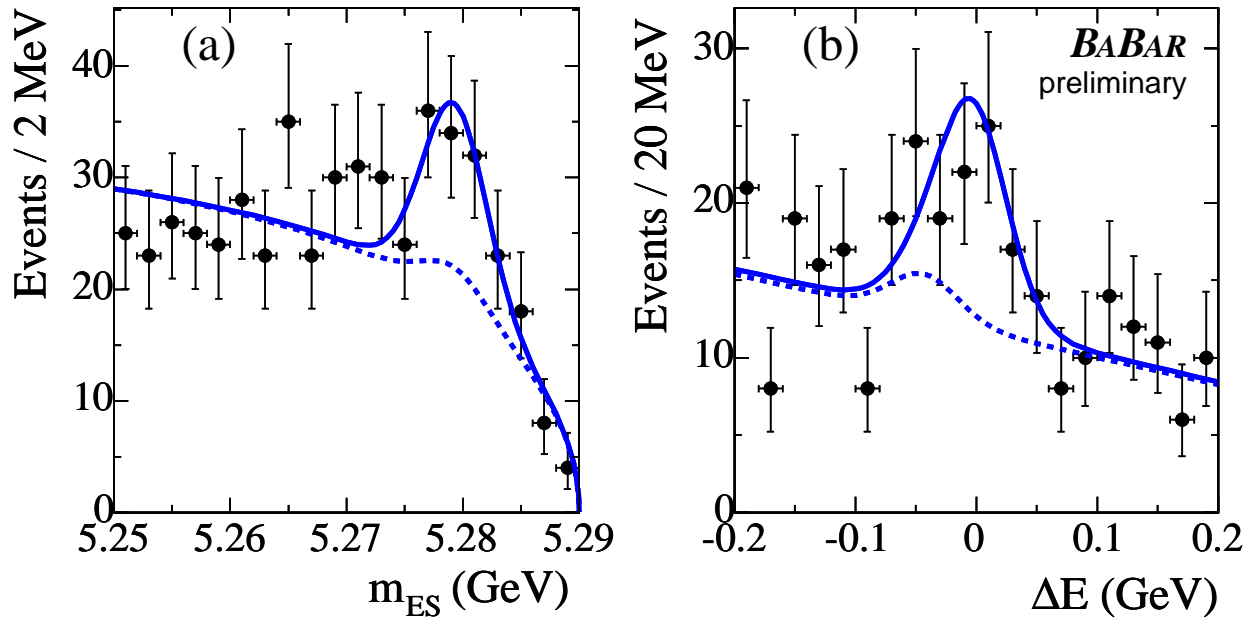}
 \caption{\label{fig:proj_deMb_etappi}
Projections of the $B$ candidate \mes\ (a) and \DE\ (b) for
\etappip.  Points with errors represent data, 
solid curves the full fit functions, 
 and dashed curves the continuum plus \etapKp\ background functions.
These plots are made with a requirement on the likelihood and thus do not 
show all events in the data sample.
  }
\end{figure}

In Figs.\ \ref{fig:proj_etaomega}--\ref{fig:proj_deMb_etappi} we show 
projections onto \mes\ and \DE\ 
of subsamples enriched with a mode-dependent threshold requirement on
the signal likelihood (computed without the PDF associated with the
variable plotted).

\begin{table*}[btp]
\caption{
Signal yield $Y$, estimated purity $P$, detection
efficiency $\epsilon$, daughter branching fraction product,
significance $S$ (with systematic uncertainties included), measured branching fraction,
background ($\acp^{qq}$) and signal (\acp) charge asymmetries for each mode.
}
\label{tab:results}
\begin{tabular}{lcccccccc}
\dbline
Mode	      	& Y	   	&$P$ &$\epsilon$ &$\prod\calB_i$ & $S$ &  \calB		& $\acp^{qq}$ & \acp\  \\
		& 			&(\%)	& (\%)	& (\%)	&$\sigma$		& $(10^{-6})$	& (\%)		& (\%)		\\
\tbline
~~\fetaggomega	&$12^{+7}_{-6}$		& 28	&13	&35	&2.4			&$1.4^{+0.7}_{-0.6}$	&   &    \\
~~\fetapppomega	&$-1^{+7}_{-5}$		& ---	&13	&20	&0.0			&$-0.2^{+1.4}_{-1.0}$	&   &    \\
\bma{\fetaomega}&                   	&	&  	&  	&\bma{\setaomega} 	&\bma{\retaomega}	&   &    \\
~~\fetaggKz	&$19^{+8}_{-7}$		& 34	&29	&14	&3.7		&$2.7^{+1.1}_{-1.0}$	&   &    \\
~~\fetapppKz	&$ 6^{+5}_{-4}$		& 30	&22	& 8	&2.1			&$1.8^{+1.6}_{-1.1}$	&   &    \\
\bma{\fetaKz}	&                   	&  	&  	&  	&\bma{\setaKz} 		&\bma{\retaKz}	&   &    \\
~~\fetaggrhop	&$110^{+31}_{-29}$	& 13  	&16	&39	&3.2   			&$ 8.1^{+2.9}_{-2.7}$	&$ 0.2\pm0.5$	&$20\pm23$    \\
~~\fetappprhop	&$ 53^{+19}_{-17}$	& 14	&11	&23	&2.8			&$ 9.7^{+4.3}_{-3.9}$	&$-0.1\pm0.8$	&$-18\pm32$    \\
\bma{\fetarhop}	&                 	&  	&  	&  	&\bma{\setarhop}	&\bma{\retarhop}	&   &\Aetarhop    \\
~~\fetapepppip	&  $55^{+12}_{-11}$	& 41	&27	&18	&4.9			&$5.4^{+1.4}_{-1.3}$	&$-0.4\pm1.4$	&$19\pm21$    \\
~~\fetaprgpip	&  $30^{+15}_{-14}$	& 14   	&18	&30	&1.2   			&$1.9^{+1.6}_{-1.4}$	&$-1.2\pm0.9$	&$47\pm44$    \\
\bma{\fetappip}	&                 	&  	&  	&  	&\bma{\setappip}	&\bma{\retappip}	&   &\Aetappip    \\
\dbline
\end{tabular}
\vspace{-5mm}
\end{table*}

\section{STATISTICAL AND SYSTEMATIC UNCERTAINTIES}
\label{sec:Errors}

The statistical error on the signal yield and \acp\ is taken as the
change in  value that corresponds to an increase of $-2\ln{\cal L}$
by one unit from its minimum.  The significance is taken as the
square root of the difference between the value of $-2\ln{\cal L}$ (with
systematic uncertainties included) for zero signal and the value at its
minimum.  For \fetaomega\ we quote a 90\% confidence level (C.L.) 
upper limit, taken to be the branching fraction below which lies 90\% of
the total of the likelihood integral in the positive branching fraction
region.  For the charged modes we also give the charge asymmetry \acp .

For the \etaph\ fits we obtain yields also for the \etapKp\ decays.  For
both submodes these yields are consistent with the expectation from our
previous measurements \cite{etaprPRL}.

Most of the systematic uncertainties arising from lack of knowledge of the
PDFs have been included in the statistical error since most background
parameters are free in the fit.  For the signal the uncertainties in PDF
parameters are estimated from the consistency of fits to MC and data in
control modes.  Varying the signal PDF parameters within these errors,
we estimate the uncertainties in the signal PDFs to be 1--8 events,
depending on the mode.  
The uncertainty in the fit bias correction is taken to be half of the
correction itself.  Similarly we estimate the uncertainty from modeling
the \BB\ backgrounds by taking half of the contribution of
that component to the fitted signal yield.
These additive systematic errors are small for the \fetaomega\ and
\fetaKz\ modes, but dominant for \fetarhop\ and \fetapip.

Uncertainties in our knowledge of the efficiency, found from auxiliary 
studies, include 0.8$N_t$\%, 1.5$N_\gamma$\%, and 3.4\%\ for a
\KS\ decay, where $N_t$ and $N_\gamma$ are the number of signal tracks
and photons, respectively.  Our estimate of the $B$ production
systematic error is 1.1\%.  Published data \cite{PDG2002}\ provide the
uncertainties in the $B$-daughter product branching fractions (1\%).
The uncertainties in the efficiency from the event selection
are 1\% (3\%\ in \etaprgpip) for the requirement on
\costhr\ and $\sim$1\% for PID.  Using several large inclusive kaon and
$B$-decay samples, we find a systematic uncertainty for \acp\ of 1.1\%,
due mainly to the dependence of the reconstruction efficiency on the charge,
for the high momentum pion from \etapip.  The corresponding number for
the softer charged pion from the $\rho$ in \etarhop\ is 2\%. The values
of $\acp^{qq}$ (see Table \ref{tab:results}) provide confirmation of
this estimate. 

\begin{figure}
\begin{center}
\scalebox{1}{
  \includegraphics[width=.5\linewidth]{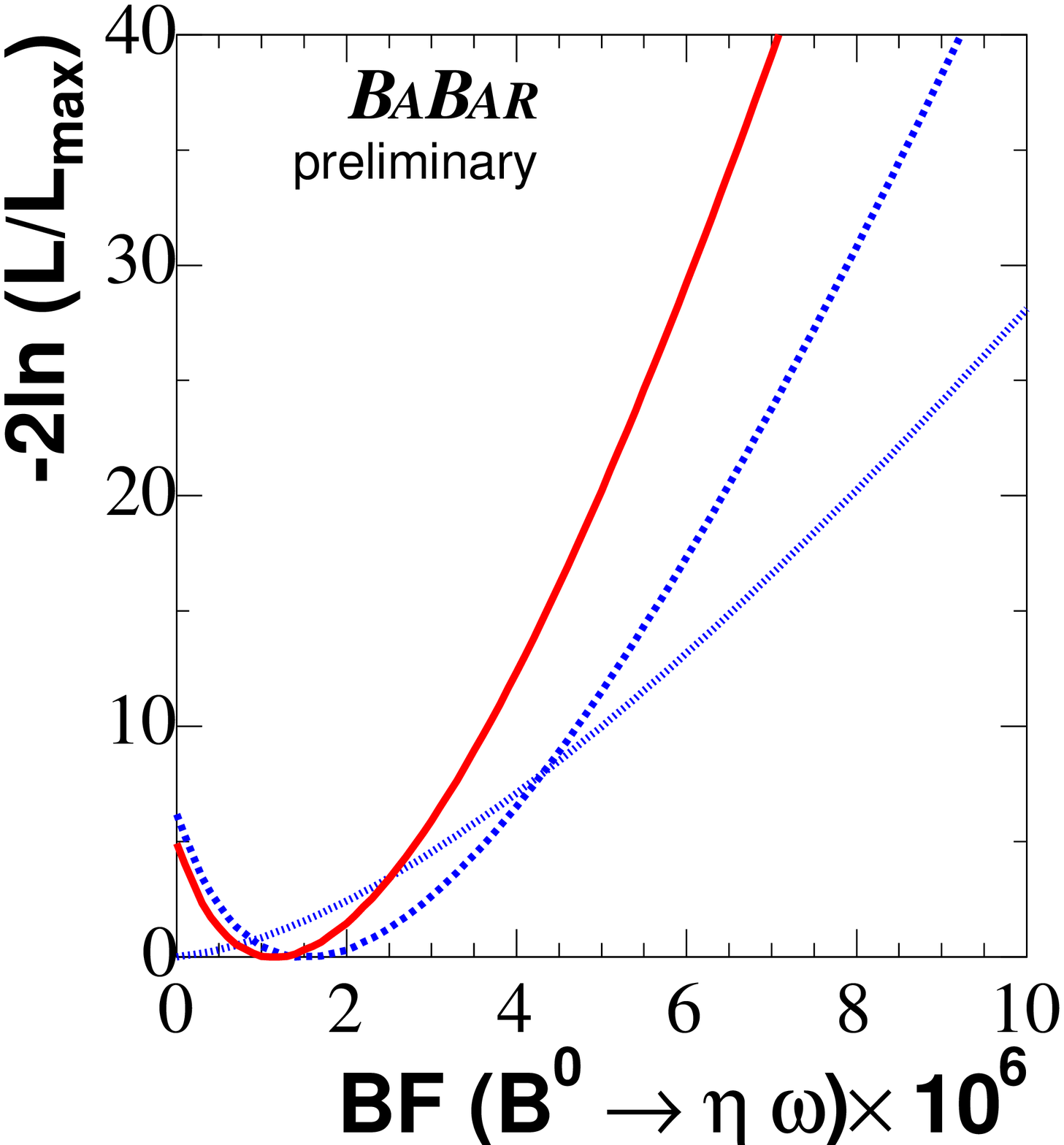}
  \includegraphics[width=.5\linewidth]{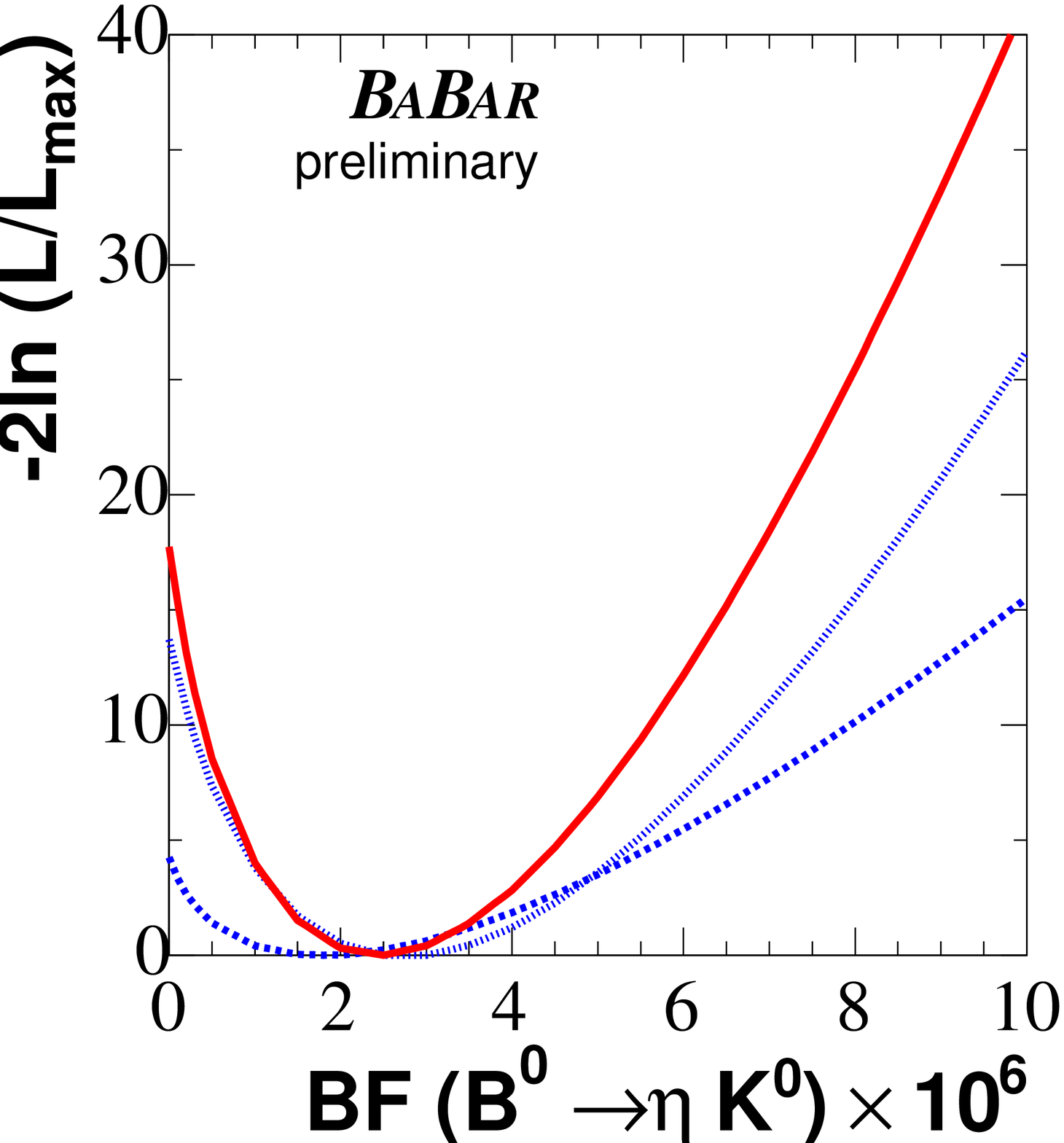}
}
\scalebox{1}{
 \includegraphics[width=1.0\linewidth]{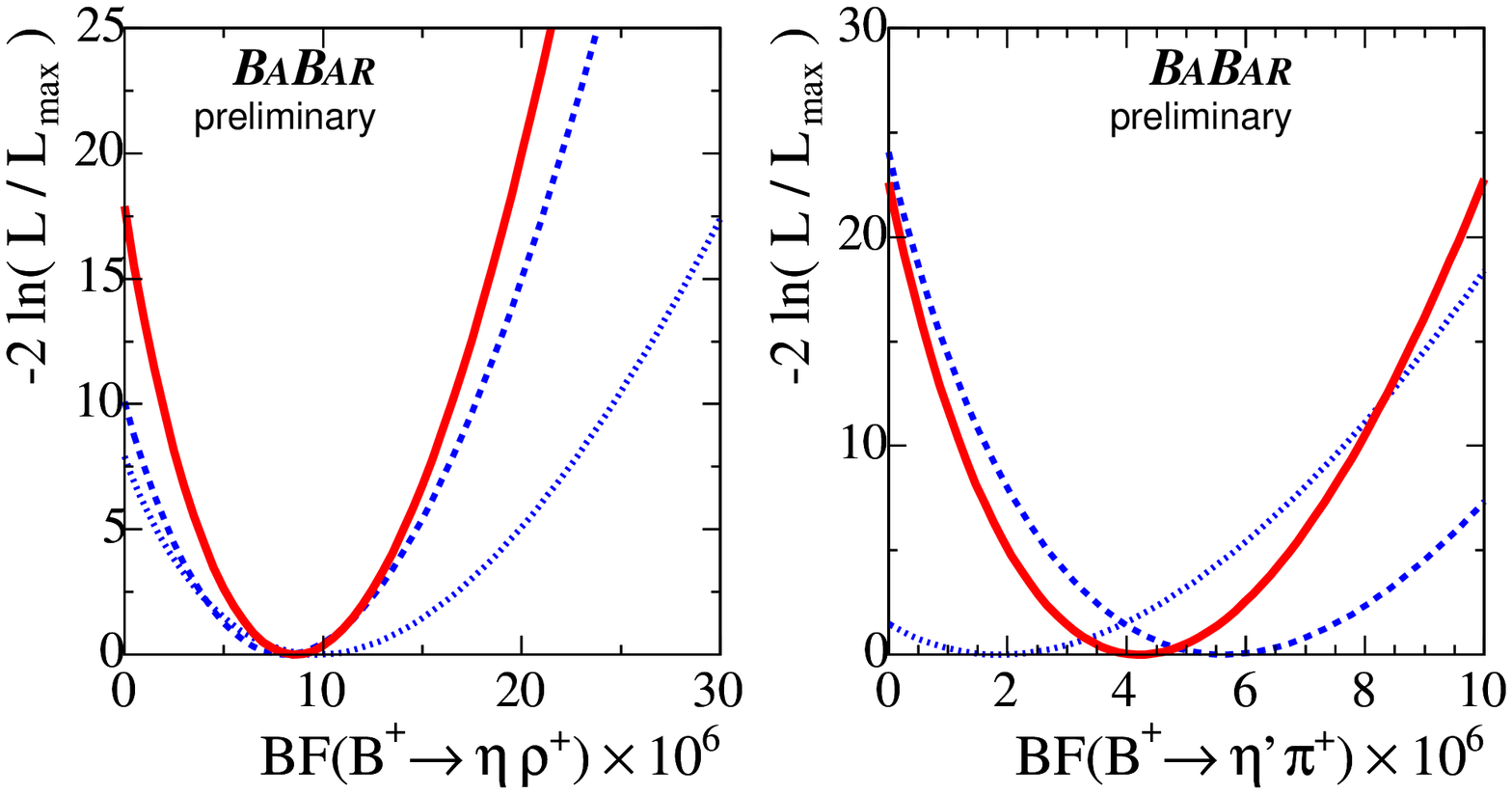}
}
\end{center}
\caption{\label{fig:combine}
Plots of individual and combined $-2\ln{\calL}$ for branching fraction
fits are shown for the decay \etaomega\ (upper left), \etaKz\ (upper
right), \etarhop\ (lower left), and \etappip\ (lower right).  Each plot
shows the daughter modes as curves that are dashed (\etaggomega,
\etapppKz, \etaggrhop, \etapepppip) or dotted (\etapppomega, \etaggKz,
\etappprhop, \etaprgpip), and the result of 
combining these as a solid curve.}
\end{figure}

The pairs of separate daughter-decay measurements for each mode are
combined by adding the values of $-2\ln{\cal L}$ as functions of
branching fraction, taking proper account
of the correlated and uncorrelated systematic errors.  We show these
curves in Fig.\ \ref{fig:combine}.

\section{SUMMARY OF RESULTS}
\label{sec:Conclusion}

In summary, we report preliminary results of searches for four
charmless $B$-meson decays.  We find significant signals for the
previously-undetected \etaKz, \etarhop, and \etappip.  The measured
branching fractions are 
\begin{eqnarray*}
\Betaomega &=& \Retaomega\quad(<\ULetaomega)\,, \\
\BetaKz &=& \RetaKz\,, \\
\Betarhop &=& \Retarhop\,, \\
\Betappip &=& \Retappip\,,
\end{eqnarray*}
where the first error quoted is statistical, the second systematic; the
upper limit is taken at 90\% CL.  For the \Bpm\ modes the charge
asymmetries, are 
\begin{eqnarray*}
\acp(\fetarhop) &=& (\Aetarhop)\%\,, \\
\acp(\fetappip) &=& (\Aetappip)\%\,.
\end{eqnarray*}

Theoretical approaches to the study of these decays include those based
on flavor SU(3) relations 
among many modes \cite{FU,FUglob,chiang,chiangGlob}, effective Hamiltonians with
factorization and specific $B$-to-light-meson form factors \cite{ALI},
perturbative QCD \cite{LEPAGE}, and QCD factorization
\cite{acpQCDfact,BENEKE,BN}. 
Our branching fraction measurements are generally in agreement with the
ranges of these theoretical estimates.  From global fits to the growing
body of data on charmless $B$ decays the component amplitudes and
theoretically uncertain parameters of these models are coming to be
significantly over-constrained \cite{FUglob,BN,chiangGlob}.
Our measurement of \acp\ in \etappip\ excludes the larger-magnitude
negative values among the theoretical estimates
\cite{acpgrabbag,acpQCDfact}. 

\section{ACKNOWLEDGMENTS}
\label{sec:Acknowledgments}


We are grateful for the 
extraordinary contributions of our \pep2\ colleagues in
achieving the excellent luminosity and machine conditions
that have made this work possible.
The success of this project also relies critically on the 
expertise and dedication of the computing organizations that 
support \babar.
The collaborating institutions wish to thank 
SLAC for its support and the kind hospitality extended to them. 
This work is supported by the
US Department of Energy
and National Science Foundation, the
Natural Sciences and Engineering Research Council (Canada),
Institute of High Energy Physics (China), the
Commissariat \`a l'Energie Atomique and
Institut National de Physique Nucl\'eaire et de Physique des Particules
(France), the
Bundesministerium f\"ur Bildung und Forschung and
Deutsche Forschungsgemeinschaft
(Germany), the
Istituto Nazionale di Fisica Nucleare (Italy),
the Foundation for Fundamental Research on Matter (The Netherlands),
the Research Council of Norway, the
Ministry of Science and Technology of the Russian Federation, and the
Particle Physics and Astronomy Research Council (United Kingdom). 
Individuals have received support from 
CONACyT (Mexico),
the A. P. Sloan Foundation, 
the Research Corporation,
and the Alexander von Humboldt Foundation.


\begin{thebibliography}{99}

\bibitem{isoscalarPRL}
\babar\ Collaboration, B. Aubert \etal, hep-ex/0403046  (to appear in
\jprl, 2004).

\bibitem{etappiPRL}
\babar\ Collaboration, B. Aubert \etal, \jprl{92}, 061801 (2004).

\bibitem{etaPRD}
\babar\ Collaboration, B. Aubert \etal,  hep-ex/0403025 (to appear in
\jprd, 2004).

\bibitem{etaprPRL}
\babar\ Collaboration, B. Aubert \etal, \jprl{91}, 161801 (2003).

\bibitem{Lipkin}
H.\ J.\ Lipkin, \plb{254}, 247 (1991).

\bibitem{directCP}
M. Bander, D. Silverman, and A. Soni, \jprl{43}, 242 (1979);
S. Barshay, D. Rein, and L.M. Sehgal, \plb{259}, 475 (1991);
A.S. Dighe, M. Gronau, and J.L. Rosner, \jprl{79}, 4333 (1997).

\bibitem{acpgrabbag}
G. Kramer, W.F. Palmer, and H. Simma, \npb{428}, 77 (1994);
A. Ali, G. Kramer, and C.-D. L\"{u}, \jprd{59}, 014005 (1999).

\bibitem{acpQCDfact}
M.-Z. Yang and Y.-D. Yang, \npb{609}, 469 (2001);
M. Beneke and M. Neubert, \npb{651}, 225 (2003).

\bibitem{FU}
H.~K.~Fu  \etal, \npps{115}, 279 (2003).

\bibitem{FUglob}
H.~K.~Fu \etal, \jprd{69}, 074002 (2004)
[hep-ph/0304242].

\bibitem{ALI}
M.~Bauer \etal, Z. Phys. C {\bf 34}, 103 (1987); 
A.~Ali and C.~Greub, \jprd{57}, 2996 (1998);
A.~Ali, G.~Kramer, and C.~D.~Lu, \jprd{58}, 094009  (1998);
Y.~H.~Chen \etal, \jprd{60}, 094014 (1999); 
J.~H.~Jang \etal, \jprd{59}, 034025 (1999).

\bibitem{LEPAGE}
G.~P.~Lepage and S.~Brodsky, \jprd{22}, 2157 (1980); 
J.~Botts and G.~Sterman, \npb{325}, 62 (1989);
Y.~Y.~Keum \etal, \plb{504}, 6  (2001), \jprd{63}, 054006  (2001);
Y.~Y.~Keum and H.~N.~Li, \jprd{63}, 074008 (2001).

\bibitem{BENEKE}
M.~Beneke \etal, \jprl{83}, 1914  (1999),
\npb{606}, 245  (2001).

\bibitem{BN}
M. Beneke and M. Neubert, \npb{675}, 333 (2003) [hep-ph/0308039] and
references therein. 

\bibitem{chiang}
C.-W. Chiang, M. Gronau, and J. L. Rosner, \jprd{68}, 074012 (2003).

\bibitem{chiangGlob}
C.-W. Chiang \etal, \jprd{69}, 034001 (2004) [hep-ph/0307395];
C.-W. Chiang \etal, hep-ph/0404073 (2004).

\bibitem{BABARNIM}
\babar\ Collaboration, B.\ Aubert \etal, \nima{479}, 1 (2002).

\bibitem{pep} 
PEP-II Conceptual Design Report, SLAC-R-418 (1993).

\bibitem{fisher} 
Fisher, R. A., Annals of Eugenics {\bf 7}, 179 (1936).

\bibitem{geant}
The \babar\ detector Monte Carlo simulation is based on GEANT4:
S. Agostinelli \etal, \nima{506}, 250 (2003).

\bibitem{PDG2002}
Particle Data Group, 
K.~Hagiwara \etal, \jprd{66}, 010001 (2002).

\end{thebibliography}
\end{document}